\documentclass[pra,final,twocolumn,showpacs]{revtex4-1}
\usepackage{graphicx,amsmath,amsfonts,wasysym,color}
\usepackage{notoccite}
\usepackage[colorlinks, linkcolor = black, citecolor = black, filecolor = black, urlcolor = blue]{hyperref}

\newcommand{\bfB}{{\bf B}}

\newcommand{\bfr}{{\bf r}}
\newcommand{\balpha}{{\mbox{\boldmath$\alpha$}}}

\newcommand{\matrixel}[3]{\langle #1 | #2 | #3 \rangle}
\newcommand{\veps}{\varepsilon}
\newcommand{\muB}{\mu_\textrm{B}}
\newcommand{\Vmagn}{V_\textrm{m}}
\newcommand{\aZ}{\alpha Z}
\newcommand{\gBB}{g_J^{(2)}}
\newcommand{\gBBB}{g_J^{(3)}}
\newcommand{\aB}{a_J^{(1)}}
\newcommand{\aBB}{a_J^{(2)}}
\newcommand{\aBBB}{a_J^{(3)}}
\newcommand{\dEFS}{\Delta E_{\textrm{FS}}}

\newcommand{\Ket}[1]{\ensuremath{\left|{#1}\right\rangle}}

\newcommand{\AngK}[2]{\Ket{{#1}/2,{#2}/2}}

\makeatletter
\AtBeginDocument{\@ifpackageloaded{natbib}{\ifNAT@numbers\if@filesw\immediate\write\@auxout{\string\global\string\NAT@numberstrue}\fi\fi}{}
}
\makeatother

\begin{document}

\title{Experimental access to higher-order Zeeman effects \\by precision spectroscopy of highly charged ions in a Penning trap}

\author{D. von Lindenfels}
\affiliation{GSI Helmholtzzentrum f\"ur Schwerionenforschung, Planckstrasse 1, 64291 Darmstadt, Germany}
\affiliation{Max-Planck-Institut f\"ur Kernphysik, Saupfercheckweg 1, 69117 Heidelberg, Germany}
\author{M. Wiesel}
\affiliation{Institut f\"ur Angewandte Physik, Technische Universit\"at Darmstadt, 64289 Darmstadt, Germany}
\affiliation{Physikalisches Institut, Im Neuenheimer Feld 226, 69120 Heidelberg, Germany}
\author{D. A. Glazov}
\affiliation{Department of Physics, St. Petersburg State University, Oulianovskaya 1, Petrodvorets, 198504 St. Petersburg, Russia}
\affiliation{Institut f\"ur Theoretische Physik, Technische Universit\"at Dresden, Mommsenstra{\ss}e 13, 01062 Dresden, Germany}
\author{A. V. Volotka}
\affiliation{Department of Physics, St. Petersburg State University, Oulianovskaya 1, Petrodvorets, 198504 St. Petersburg, Russia}
\affiliation{Institut f\"ur Theoretische Physik, Technische Universit\"at Dresden, Mommsenstra{\ss}e 13, 01062 Dresden, Germany}
\author{M. M. Sokolov}
\affiliation{Department of Physics, St. Petersburg State University, Oulianovskaya 1, Petrodvorets, 198504 St. Petersburg, Russia}
\author{G. Plunien}
\affiliation{Institut f\"ur Theoretische Physik, Technische Universit\"at Dresden, Mommsenstra{\ss}e 13, 01062 Dresden, Germany}
\author{V. M. Shabaev}
\affiliation{Department of Physics, St. Petersburg State University, Oulianovskaya 1, Petrodvorets, 198504 St. Petersburg, Russia}
\author{W. Quint}
\affiliation{GSI Helmholtzzentrum f\"ur Schwerionenforschung, Planckstrasse 1, 64291 Darmstadt, Germany}
\affiliation{Physikalisches Institut, Im Neuenheimer Feld 226, 69120 Heidelberg, Germany}
\author{G. Birkl}
\affiliation{Institut f\"ur Angewandte Physik, Technische Universit\"at Darmstadt, 64289 Darmstadt, Germany}
\author{A. Martin}
\affiliation{Institut f\"ur Angewandte Physik, Technische Universit\"at Darmstadt, 64289 Darmstadt, Germany}
\author{M. Vogel}
\email{m.vogel@gsi.de}
\affiliation{Institut f\"ur Angewandte Physik, Technische Universit\"at Darmstadt, 64289 Darmstadt, Germany}


\begin{abstract}
We present an experimental concept and setup for laser-microwave double-resonance spectroscopy of
highly charged ions in a Penning trap. Such spectroscopy allows a highly precise measurement of the
Zeeman splittings of fine- and hyperfine-structure levels due the magnetic field of the trap. We have
performed detailed calculations of the Zeeman effect in the framework of quantum electrodynamics of
bound states as present in such highly charged ions. We find that apart from the linear Zeeman effect,
second- and third-order Zeeman effects also contribute to the splittings on a level of $10^{-4}$ and
$10^{-8}$, respectively, and hence are accessible to a determination within
the achievable spectroscopic resolution of the ARTEMIS experiment currently in preparation.
\end{abstract}


\maketitle

\section{Introduction}
Ever since the discovery of a quadratic contribution to the Zeeman effect by Jenkins and Segr\'e in
the 1930s \cite{zee1,zee2}, there have been numerous studies both experimental and theoretical of
higher-order Zeeman contributions in atoms, molecules, and singly charged ions in laboratory magnetic
fields (see, for example, \cite{zee2b,zee2c,zee2d,zee3}). The high magnetic field strengths present in
astronomical objects have given impetus to corresponding studies in observational astronomy
\cite{pre,ham,kem,as1,dwa}, identifying a quadratic Zeeman effect in abundant species like hydrogen
and helium.
Although highly charged ions are both abundant in the universe and readily accessible in laboratories,
to our knowledge, no higher-order Zeeman effect in highly charged ions has been observed so far.

In highly charged ions of a given charge state, the electronic energy level splittings depend strongly
on the nuclear charge $Z$. For one-electron ions (i.e., hydrogenlike ions) the energy splitting is
proportional to $Z^2$ for principal transitions, to $Z^3$ for hyperfine-structure transitions, and to
$Z^4$ for fine-structure transitions. In other few-electron ions the scaling is very similar
\cite{bey1,bey2}. Since in the hydrogen atom
principal transitions are typically at a few eV,
the scaling with $Z^2$ shifts these transitions far into the XUV and x-ray regime for heavier
hydrogenlike ions, and thus out of the reach of studies like the present one.

In an external magnetic field, the Zeeman effect lifts the degeneracy of energies within fine- and
hyperfine-structure levels. For highly charged ions in magnetic fields of a few tesla strength as
typical for Penning trap operation, the corresponding Zeeman splitting is well within the microwave
domain and thus accessible for precision spectroscopy. In addition, in fine- and
hyperfine-structure transitions, the strong scaling with $Z$ eventually shifts the corresponding
energies into the laser-accessible region and thus makes them available for precision optical
spectroscopy \cite{vogpr}.

We are currently setting up a laser-microwave double-resonance spectroscopy experiment with highly
charged ions in a Penning trap, which combines precise spectroscopy of both optical transitions and
microwave Zeeman splittings \cite{quintpra,dav}. The experiment aims at spectroscopic precision
measurements of such energy level splittings and magnetic moments of bound electrons on the ppb
level of accuracy and better. At the same time, it allows access to the nuclear magnetic moment
in the absence of diamagnetic shielding \cite{quintpra,yerokhin:11:prl}. For first tests within the
AsymmetRic Trap for the measurement of Electron Magnetic moments in IonS (ARTEMIS)
experiment, the $^{40}$Ar$^{13+}$ (spectroscopic notation: ArXIV) ion has been chosen.
It has a spinless nucleus, such
that only a fine structure is present. Similar measurements in hyperfine structures are to be
performed with ions of higher charge states such as, for example, $^{207}$Pb$^{81+}$ and
$^{209}$Bi$^{82+}$ as available to ARTEMIS within the framework of the HITRAP facility
\cite{kluge} at GSI, Germany.

We have performed detailed relativistic calculations of the Zeeman effect in boronlike ions such
as Ar$^{13+}$. These calculations show that at the ppb level of experimental accuracy, higher-order
effects play a significant role and need to be accounted for. In turn, precision spectroscopy of
highly charged ions allows a measurement of these higher-order contributions to the Zeeman effect.

\section{Calculation of the Zeeman Effect}\label{sec:theo}
\begin{figure}[h!tb]
\begin{center}
\includegraphics[width=0.52\textwidth]{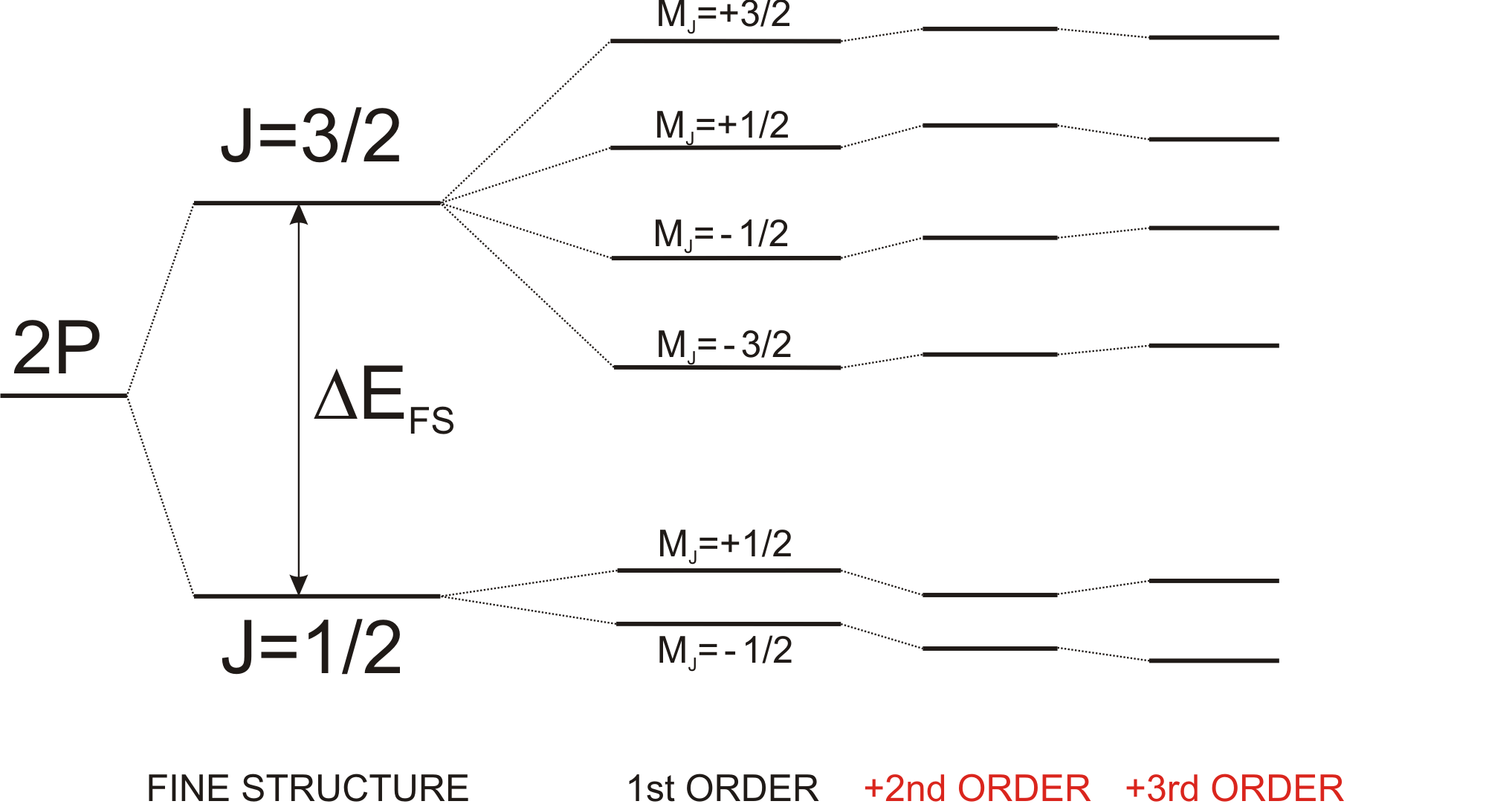}
\caption{\small (Color online) Level scheme of the $2\,{}^2\!P_J$ states of boronlike argon ArXIV in an external
magnetic field with higher-order contributions to the Zeeman effect (not true to scale).}
\label{fig:levels}
\end{center}
\end{figure}

We consider a five-electron argon ion in the ground $[(1s)^2(2s)^22p]\,{}^2\!P_{1/2}$ and in the
first excited $[(1s)^2(2s)^22p]\,{}^2\!P_{3/2}$ states. The fine-structure interval between these
levels has previously been studied \cite{draganic:03:prl,artemyev:07:prl,soriaorts:07:pra,maeckel:11:prl},
as well as the corresponding magnetic dipole transition rate
\cite{tupitsyn:05:pra,lapierre:05:prl,volotka:06:epjd,volotka:08:epjd}. An external magnetic field
splits levels with different angular momentum projection onto the direction of the field. While
this splitting is equidistant in the first-order approximation, the nonlinear magnetic field
effects disturb this symmetry. The corresponding level structure is schematically depicted in
Fig. \ref{fig:levels}.

The Zeeman shift of each level can be evaluated within perturbation theory,
\begin{eqnarray}
E_A(B) &=& E_A^{(0)} + \Delta E_A^{(1)}(B)
\nonumber\\
&& + \Delta E_A^{(2)}(B) + \Delta E_A^{(3)}(B) + \cdots 
,
\end{eqnarray}
where $\Ket{A}=\Ket{J,M_J}$ is the $2\,{}^2\!P_J$ state with total angular momentum $J$ and its
projection $M_J$. In the following, the quantities which do not depend on $M_J$ (e.g., the energy in
the absence of magnetic field, $E_A^{(0)}$) are labeled with $J$ only. Each term of the perturbation
expansion is proportional to the magnetic field strength to the corresponding power,
$\Delta E_A^{(n)}(B) \sim B^n$. The first-order term is directly related to the $g_J$ factor by
\begin{equation}
\label{eq:g}
\Delta E_A^{(1)}(B) = g_J M_J \muB B
,
\end{equation}
where $\muB$ is the Bohr magneton.
The Dirac equation for the valence $\Ket{a}=\Ket{2p}$ electron is an appropriate zeroth approximation
to find $\Delta E_A^{(1)}(B)$ as
\begin{equation}
\label{eq:aVa}
\Delta E_A^{(1)}(B) = \matrixel{a}{V_\textrm{m}}{a}
,
\end{equation}
where the operator
\begin{equation}
\Vmagn = \frac{|e|}{2}\, [\bfr\times\balpha] \cdot \bfB
\end{equation}
represents the interaction with the external homogeneous magnetic field $\bfB$.
For the Coulomb potential of a pointlike nucleus, one finds
\begin{eqnarray}
g_{1/2} &=& \frac{2}{3} \left[ \sqrt{ 2 \left [ 1 + \sqrt{1-(\aZ)^2} \right] } - 1 \right] 
\nonumber\\
&=& \frac{2}{3} - \frac{1}{6} (\aZ)^2 - \frac{5}{96} (\aZ)^4 - \cdots
,\\
g_{3/2} &=& \frac{4}{15} \left[ 2 \sqrt{ 4 - (\aZ)^2 } + 1 \right] 
\nonumber\\
&=& \frac{4}{3} - \frac{2}{15} (\aZ)^2 - \frac{1}{120} (\aZ)^4 - \cdots 
.
\end{eqnarray}
The interelectronic interaction, quantum electrodynamical, and nuclear effects give rise to corrections
to these values.
Evaluation of the $g_J$ factors of the $2\,{}^2\!P_{1/2}$ and $2\,{}^2\!P_{3/2}$ states of boronlike
argon in Ref. \cite{soriaorts:07:pra} yielded $g_{1/2}=0.663\,65$ and $g_{3/2}=1.332\,28$. These values
include the one-loop QED term and the interelectronic interaction correction. The latter was calculated
within the configuration-interaction method with the basis functions derived from the Dirac-Fock and
Dirac-Fock-Sturm equations. The contribution of the negative-energy states, which is crucially important
for the Zeeman effect, was taken into account within perturbation theory.
Recently, the $g_J$ factors have been improved to $g_{1/2}=0.663\,647(1)$ and $g_{3/2}=1.332\,285(3)$
\cite{glazov:ps}. In comparison to those from Ref. \cite{soriaorts:07:pra}, these include the
$1/Z$ term of the interelectronic interaction, evaluated within the QED approach, the screening
correction to the one-loop self-energy term, and the nuclear recoil effect.

The second- and third-order terms in the Zeeman splitting can be presented in the following form:
\begin{eqnarray}
\label{eq:gBB}
\Delta E_A^{(2)}(B) = \gBB(M_J) (\muB B)^2 / E_0
,
\\
\label{eq:gBBB}
\Delta E_A^{(3)}(B) = \gBBB(M_J) (\muB B)^3 / E_0^2
,
\end{eqnarray}
where $E_0 = mc^2$ is the electron rest energy, while $\gBB$ and $\gBBB$ are dimensionless coefficients.
Their dependence on $M_J$ is not as simple as for the first-order effect; however, they obey the symmetry
relations $\gBB(-M_J)=\gBB(M_J)$ and $\gBBB(-M_J)=-\gBBB(M_J)$.

The leading-order contributions to $\Delta E_A^{(2)}$ and $\Delta E_A^{(3)}$ can be calculated according
to the formulas
\begin{eqnarray}
\label{eq:aVVa}
\Delta E_A^{(2)}(B) &=& {\sum_{n}}' \frac{\matrixel{a}{\Vmagn}{n}\matrixel{n}{\Vmagn}{a}}{\veps_a - \veps_n}
,\\
\Delta E_A^{(3)}(B) &=& {\sum_{n_1,n_2}}' \frac{\matrixel{a}{\Vmagn}{n_1}\matrixel{n_1}{\Vmagn}{n_2}\matrixel{n_2}{\Vmagn}{a}} {(\veps_a-\veps_{n_1})(\veps_a-\veps_{n_2})}
\nonumber\\
\label{eq:aVVVa}
&-& {\sum_{n}}' \frac{\matrixel{a}{\Vmagn}{n}\matrixel{n}{\Vmagn}{a}}{(\veps_a - \veps_n)^2} \matrixel{a}{\Vmagn}{a}
,
\end{eqnarray}
where the summations run over the complete Dirac spectrum, excluding the reference state $\Ket{a}$. It
is absolutely important to take into account the negative-energy states in Eqs. (\ref{eq:aVVa}) and
(\ref{eq:aVVVa}), since their contribution is not small as compared to that of the positive-energy
states even for low nuclear charge $Z$ (nonrelativistic limit). In particular, in hydrogenlike
($\Ket{a}=\Ket{1s}$) and lithiumlike ($\Ket{a}=\Ket{2s}$) ions the negative continuum delivers
a dominant part of these higher-order terms. However, in magnetic fields of several tesla, their
magnitudes appear to be far below the experimental precision. For example, the $g_J$ factors of
hydrogen- and lithiumlike silicon ions have been measured recently with ppb accuracy at a
magnetic field of 3.76 T \cite{sturm:11:prl,wagner:prl}. In these cases, the relative contribution
of the third-order effect $|\Delta E^{(3)}/\Delta E^{(1)}|$ is $0.05 \cdot 10^{-15}$ and
$0.8 \cdot 10^{-15}$, respectively. Although the quadratic shift is not so small,
$|\Delta E^{(2)}/\Delta E^{(1)}| = 0.7 \cdot 10^{-6}$ for lithiumlike silicon, it does not
affect the ground-state Zeeman splitting for states with $J=1/2$. Just as in the present case
(see Fig. \ref{fig:levels}), both sublevels are shifted by the same amount,
which cancels in the transition frequency.

In contrast, in boronlike ions the higher-order effects appear to be well observable. This is due to
the relatively small fine-structure interval between the states $2p_{1/2}$ and $2p_{3/2}$. Below we
consider the Zeeman shifts for both of these states. While $\Ket{a}=\Ket{2p_{1/2}}$,$\Ket{2p_{3/2}}$
denotes the reference state, $\Ket{b}=\Ket{2p_{3/2}},\Ket{2p_{1/2}}$ denotes the other state of these two.
It has been verified by rigorous calculations that the contribution of the fine-structure partner $\Ket{b}$
in Eqs. (\ref{eq:aVVa}) and (\ref{eq:aVVVa}) is dominant in the case of $M_J=\pm1/2$. Accordingly,
the summations can be restricted to $\Ket{n}=\Ket{n_1}=\Ket{n_2}=\Ket{b}$ to yield the estimations
\begin{eqnarray}
\label{eq:estVV}
\Delta E_A^{(2)}(B) &\approx& \frac{|\matrixel{a}{\Vmagn}{b}|^2}{\veps_a - \veps_b}
,\\
\label{eq:estVVV}
\Delta E_A^{(3)}(B) &\approx& \frac{|\matrixel{a}{\Vmagn}{b}|^2}{(\veps_a-\veps_b)^2} \left( \matrixel{b}{\Vmagn}{b} - \matrixel{a}{\Vmagn}{a} \right)
.
\end{eqnarray}
Equation (\ref{eq:estVV}) shows that $\Delta E_A^{(2)}$ is approximately of the same magnitude and of
opposite sign for the two considered states.
Equation (\ref{eq:estVVV}) shows that the same holds for $\Delta E_A^{(3)}$.
Even more simplified order-of-magnitude estimations of these effects are valid in the present case,
\begin{eqnarray}
\label{eq:omVV}
\frac{\Delta E_A^{(2)}(B)}{\Delta E_A^{(1)}(B)} &\sim& \frac{\Delta E_A^{(1)}(B)}{\dEFS} \sim 10^{-4}
,\\
\label{eq:omVVV}
\frac{\Delta E_A^{(3)}(B)}{\Delta E_A^{(1)}(B)} &\sim& \left( \frac{\Delta E_A^{(1)}(B)}{\dEFS} \right)^2 \sim 10^{-8}
,
\end{eqnarray}
where $\dEFS = E_{3/2}^{(0)}-E_{1/2}^{(0)}$ is the fine-structure interval.
Please note, however, that Eqs. (\ref{eq:estVV})--(\ref{eq:omVVV}) are justified by rigorous calculations according to Eqs. (\ref{eq:aVVa}) and (\ref{eq:aVVVa}) for the $2p_{1/2}$ and $2p_{3/2}$
states and are not necessarily valid in other cases.

We have performed the calculations according to Eqs. (\ref{eq:aVVa}) and (\ref{eq:aVVVa}) within the
dual-kinetic-balance (DKB) approach \cite{shabaev:04:prl} with the basis functions constructed from
$B$ splines \cite{sapirstein:96:jpb}. Several effective screening potentials, which partly take into
account the interelectronic-interaction effects (see, e.g.,
\cite{cowan,sapirstein:02:pra,glazov:06:pla,volotka:08:pra}), have been employed to estimate the
uncertainty of the results. As one can see from Eqs. (\ref{eq:aVVa}) and (\ref{eq:aVVVa}) the
higher-order effects are highly sensitive to the fine-structure energy splitting $\dEFS$, which
is significantly affected by the interelectronic-interaction and QED effects. Therefore, instead
of the value of $\dEFS$ provided by the Dirac equation with the screening potential we employed
the best up-to-date theoretical value from Ref. \cite{artemyev:07:prl}, which is in perfect agreement
with the experimental one \cite{maeckel:11:prl}. Finally, the values for different screening potentials
have been averaged. The results for $\gBB$ and $\gBBB$ are presented in Table \ref{tab:gBB}. They are
in agreement with the values obtained by Tupitsyn within the large-scale configuration-interaction
Dirac-Fock-Sturm method \cite{tupitsyn:unp}. We estimate the uncertainty of the values obtained roughly as $10\%$. Rigorous evaluation of the correlation effects beyond the screening-potential
approximation is needed. QED and nuclear recoil effects have to be taken into account as well.

The energies of the Zeeman sublevels including the linear and nonlinear effects can be written as
\begin{eqnarray}
E_A(B) &=& E_{{J}}^{(0)} + h \sum_{i=1}^\infty a_J^{(i)} (M_J) B^i
.
\end{eqnarray}
The coefficients $a^{(i)}_J(M_J)$ are directly related to the $g_J$, $\gBB$, and $\gBBB$ factors,
defined by Eqs. (\ref{eq:g}), (\ref{eq:gBB}), and (\ref{eq:gBBB}),
\begin{eqnarray}
\label{eq:aB}
h \aB(M_J) &=& g_J M_J \muB
,\\
\label{eq:aBB}
h \aBB(M_J) &=& \gBB(M_J) \muB^2 /(m c^2)
,\\
\label{eq:aBBB}
h \aBBB(M_J) &=& \gBBB(M_J) \muB^3 /(m c^2)^2
.
\end{eqnarray}
The values of the coefficients $\aBB$ and $\aBBB$ are presented in Table \ref{tab:gBB} along with
$\gBB$ and $\gBBB$.

%
%
\begin{table}[t]
\caption{$\gBB$ and $\gBBB$ factors [Eqs. (\ref{eq:gBB}) and (\ref{eq:gBBB})] and the corresponding
coefficients $\aBB$ and $\aBBB$
[Eqs. (\ref{eq:aBB}) and (\ref{eq:aBBB})] for boronlike argon.
\label{tab:gBB}}
\begin{tabular}{cr@{}lr@{}lll}
\hline
\hline

$J$, $M_J$
& \multicolumn{2}{c}{$\gBB(M_J)$}
& \multicolumn{2}{l}{$\aBB(M_J)$}

& $\gBBB(M_J)$
& $\aBBB(M_J)$
\\
& & &
\multicolumn{2}{l}{(kHz/T$^2$)}
& &
(Hz/T$^3$)
\\
\hline
$3/2$, $\pm 3/2$ \qquad & $0$&$.95 \cdot 10^{3}$ \qquad & $1$&$.5$ \qquad & $\mp 5.7 \cdot 10^{3}$ \qquad & $\mp 1.0 \cdot 10^{-6}$ \\
$3/2$, $\pm 1/2$ \qquad & $41$&$.0 \cdot 10^{3}$ \qquad & $65$&$.1$ \qquad & $\mp 2.5 \cdot 10^{9}$ \qquad & $\mp 0.45$ \\
$1/2$, $\pm 1/2$ \qquad & $-39$&$.5 \cdot 10^{3}$ \qquad & $-62$&$.6$ \qquad & $\pm 2.5 \cdot 10^{9}$ \qquad & $\pm 0.45$ \\
\hline
\hline
\end{tabular}
\end{table}
%
%
\begin{table}[t]
\caption{Contributions to the Zeeman energy shifts for boronlike argon at 7 T.
First, second, and third orders in the magnetic field are presented in terms
of the frequencies $\Delta E / h$. \label{tab:shifts}}
\begin{tabular}{cr@{}lr@{}lr@{}l}
\hline
\hline
$J$, $M_J$
& \multicolumn{2}{c}{$\Delta E_A^{(1)}/h$ (GHz)}
& \multicolumn{2}{c}{$\Delta E_A^{(2)}/h$ (MHz)}
& \multicolumn{2}{c}{$\Delta E_A^{(3)}/h$ (Hz)}
\\ \hline
$3/2$, $+3/2$ & 195.&793 & 0.&074 & $-$0.&00035 \\
$3/2$, $+1/2$ & 65.&264 & 3.&19 & $-$153& \\
$3/2$, $-1/2$ & $-$65.&264 & 3.&19 & 153& \\
$3/2$, $-3/2$ & $-$195.&793 & 0.&074 & 0.&00035 \\

$1/2$, $+1/2$ & 32.&5100 & $-$3.&07 & 153& \\
$1/2$, $-1/2$ & $-$32.&5100 & $-$3.&07 & $-$153& \\
\hline
\hline
\end{tabular}
\end{table}

Table \ref{tab:shifts} shows the first-, second-, and third-order contributions to the Zeeman shift of
individual levels in boronlike argon in a magnetic field of 7 T. The linear effect separates the
two ground-state ($2\,{}^2\!P_{1/2}$) levels by about 65 GHz and the four excited-state
($2\,{}^2\!P_{3/2}$) levels by about 130 GHz. The quadratic effect shifts both $\Ket{1/2,\pm 1/2}$
levels down and the two $\Ket{3/2,\pm 1/2}$ levels up by about 3 MHz. This effect is exactly independent
of the sign of $M_J$, while the shifts for $J=1/2$ and $J=3/2$ are slightly different. The
$\Ket{3/2,\pm 3/2}$ levels are shifted up by 74 kHz. So the second-order effect contributes to the
transition frequencies $\nu_a$ and $\nu_c$, which are introduced in the next section (see also Fig.
\ref{fig:spec0}). The cubic effect increases the splitting between the ground-state levels by about
306 Hz, thus simulating a contribution of $3\cdot 10^{-9}$ to $g_{1/2}$. The splitting between
the $\Ket{3/2,\pm 1/2}$ levels is decreased by approximately the same value.

\section{Double-Resonance Spectroscopy}
\subsection{General concept} \label{sec:general}
\begin{figure}[h]
\centering
\includegraphics[width=8cm]{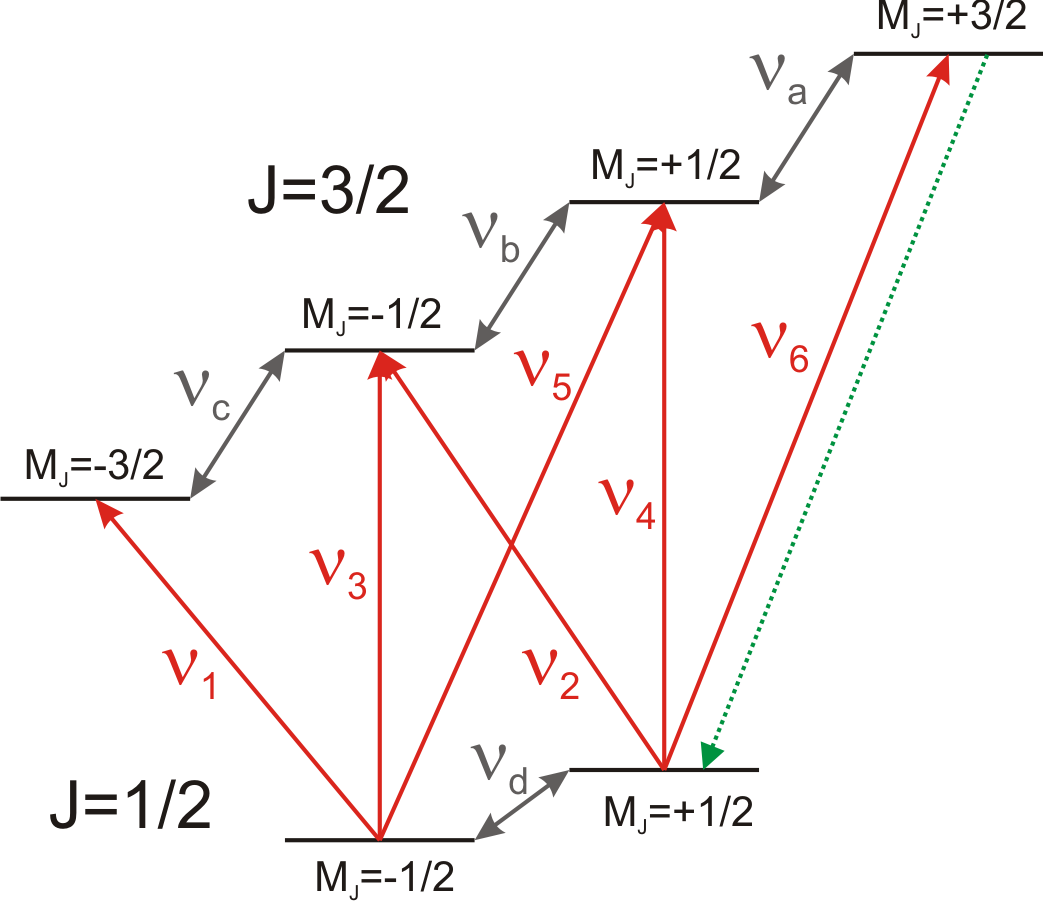}
\caption[Fine-structure level scheme of boron-like argon]{(Color online)
Spectroscopy of the $2\,^2\!P_{1/2}$-$2\,^2\!P_{3/2}$ fine-structure
transition, as in a boronlike argon ion ArXIV, with Zeeman effect. The level scheme
(not true to scale) and all magnetic dipole transitions are shown.
Solid arrows indicate excitation by laser photons,
while dotted arrows are spontaneous decays.
Gray double arrows represent microwave transitions.
\label{fig:spec0}}
\end{figure}

The technique of laser-microwave double-resonance spectroscopy has been applied in experiments with
singly charged ions \cite{lm1,lm2,lm3}.
Presently, the concept can be exemplified by Fig. \ref{fig:spec0}. It shows all six components of the
Zeeman-split magnetic dipole transition
between the two fine-structure states of a \textit{P} electron. Neighboring lines are all separated by
about the same frequency difference, because the upper-state level spacing is about twice as
large as that in the lower state.
If an ion gets excited by the most redshifted frequencies ($\nu_1$ or $\nu_2$), the angular
momentum projection is lowered by 1. The fluorescence light corresponding to these transitions
is emitted mainly along the quantization axis with circular polarization. Yet a fraction
can also be detected under radial observation. This component is polarized parallel to the axis.
The frequencies $\nu_3$ and $\nu_4$ are the closest to the field-free frequency. The corresponding
transitions preserve the angular momentum projection and can be observed only radially and with
perpendicular polarization.
Excitation with the most blueshifted lines at $\nu_5$ or $\nu_6$ increases the angular momentum
projection.
The emitted fluorescence follows the same distribution
as in the case of $\nu_1$ or $\nu_2$, but has opposite helicity.

The basic idea of double-resonance spectroscopy is contained in the following example: A closed optical
cycle between the extremal states $\Ket{1/2,+1/2}$ and $\Ket{3/2,+3/2}$ is driven resonantly by a laser
at frequency $\nu_6$. The corresponding fluorescence light (dotted arrow) is observed continuously. In
the absence of microwave radiation the fluorescence light intensity is constant. When either of the microwave
transitions at $\nu_a$ or $\nu_d$ is driven resonantly, population is withdrawn from the optical cycle
and the amount of fluorescence light is reduced. Hence, the optical signal indicates when the desired
microwave transition is resonantly driven, yielding the desired Zeeman transition frequency. A detailed
discussion of the applicability of this concept in different level-scheme situations is given in
\cite{quintpra}. Note that all Zeeman sublevels can be addressed individually since their separation
is much larger than the laser width and the Doppler widths of the optical transitions.

\subsection{Application to an optical \textit{P} doublet}
In the present application, the optical transition is a ground-state fine-structure transition in a
highly charged ion, while the
microwave transition occurs between corresponding Zeeman sub-levels.
Both are magnetic dipole (\textit{M}1) transitions with accordingly long lifetimes of the upper levels.
By laser-microwave double-resonance spectroscopy, the Zeeman sublevel splittings can be measured with
high accuracy. This yields access to differences of the coefficients defined in Sec. \ref{sec:theo}
rather than to the values themselves. Therefore we abbreviate:
\begin{align}
a_i &\equiv a_{1/2}^{(i)}(1/2) - a_{1/2}^{(i)}(-1/2),\;\;\ i=1,3, \nonumber\\
b_i &\equiv a_{3/2}^{(i)}(3/2) - a_{3/2}^{(i)}(1/2),\;\;\ i=1,2,3, \nonumber\\
b'_3 &\equiv a_{3/2}^{(3)}(3/2). \nonumber
\end{align}
Due to the tiny value of
$b'_3 B^3 < 1\;\rm{mHz}$, we cannot distinguish $b_3$ from $- a_{3/2}^{(3)}(1/2)$.

The upper-level Zeeman splitting is of particular interest, because here the quadratic shift
actually has a measurable effect on the Larmor frequencies. We name the three frequencies
$\nu_a,\ \nu_b,$ and $ \nu_c$,
while the ground-state Larmor frequency is $\nu_d$, as denoted in Fig. \ref{fig:spec0}.

The combination of particular frequencies can be used to disentangle the different orders:
\begin{align}
\nu_d &=\,\ a_1 B \quad\qquad\, + \,\ a_3 B^3, \nonumber\\
\nu_a + \nu_c &= 2 b_1 B \quad\qquad\ + 2 b_3 B^3, \nonumber\\
\nu_a - \nu_c &= \quad\qquad 2 b_2 B^2, \nonumber\\
\nu_a + \nu_b + \nu_c &= 3 b_1 B \quad\qquad\ + 2 b'_3 B^3, \nonumber\\
\nu_a + \nu_c - 2 \nu_b &= \qquad\qquad\qquad\quad\! 6 b_3 B^3 - 4 b'_3 B^3. \nonumber
\end{align}
Neglecting $b'_3 B^3$, we can immediately derive $b_1, b_2, b_3$ from the latter equations, provided
the magnetic
field strength has been measured with corresponding accuracy. In the lower state,
the quadratic effect cancels. This is different for the cubic order: So far, we have to use the theoretical
prediction for $a_3$ in
order to determine $a_1$. A similar procedure would yield $b_1$ in the case of an insufficient $\nu_b$
measurement:
\begin{align}
g_{1/2} \frac{\muB}{h} \equiv a_1
&= \frac1B \left(\quad\,\ \nu_d \quad\, - a_3 B^3 \right), \nonumber\\
g_{3/2} \frac{\muB}{h} \equiv b_1
&= \frac1B \left( \frac{\nu_a + \nu_c}{2} - b_3 B^3 \right). \nonumber
\end{align}

Under these conditions, we can imagine several different spectroscopic options. All involve blue laser
radiation ($\lambda \approx 441$ nm) and a microwave field with at least one of the aforementioned
frequencies. Therefore, the
term ``double resonance'' may be extended to ``triple'' or even ``quadruple'' resonance. In any case,
the optical spectroscopy serves to prepare or detect population in specific Zeeman states and thus
allows us to see whether the microwave frequency is at resonance with the transition of interest. The set
frequency together with the measured response of the ions
is used to determine the Larmor frequency. In the following, we will
explain the most useful and viable concept and just briefly mention variations.

For any microwave scan, we start with a spectrally broad signal (statistical or a Landau-Zener sweep)
and iteratively lower the width. This can be performed by an external modulation of the output frequency of the microwave generator.
In addition, the modulation technique allows us to apply multiple sharp frequencies at once.

The ratio of $g_J$ values in the respective states is close to 2, namely, 2.007.
The frequencies to be generated before multiplication are, according to the current theoretical estimation,
\begin{align}
\nu_a/8 &= {16.323}\; {\text{GHz}}, \nonumber\\
\nu_b/8 &= {16.323}\; \mbox{GHz}, \nonumber\\
\nu_c/8 &= {16.324}\; \mbox{GHz}, \nonumber\\
\nu_d/4 &= {16.262}\; \mbox{GHz}. \nonumber
\end{align}

These frequencies are well within the few-percent spectral acceptance of an active quadrupler
for $\nu_d$. We use an additional doubling
stage for the upper-level Larmor frequencies $\nu_a$, $\nu_b$, and $\nu_c$.
Thus we can address several upper-level Larmor transitions simultaneously, namely, by nonlinear mixing of
close-lying fundamental frequencies.

The relatively long lifetime of 9.6 ms
has two benefits: First, it
allows a
temporal separation of excitation and detection, and, second, the microwaves have enough time to stimulate transitions between the excited-level substates
in the upper fine-structure level. Anyway, the precision of the upper-level
Larmor frequencies is limited by the natural linewidth of about 100 Hz.

\subsubsection{Separation of the linear and quadratic effects}
\begin{figure}[h]
\centering
\includegraphics[width=6cm]{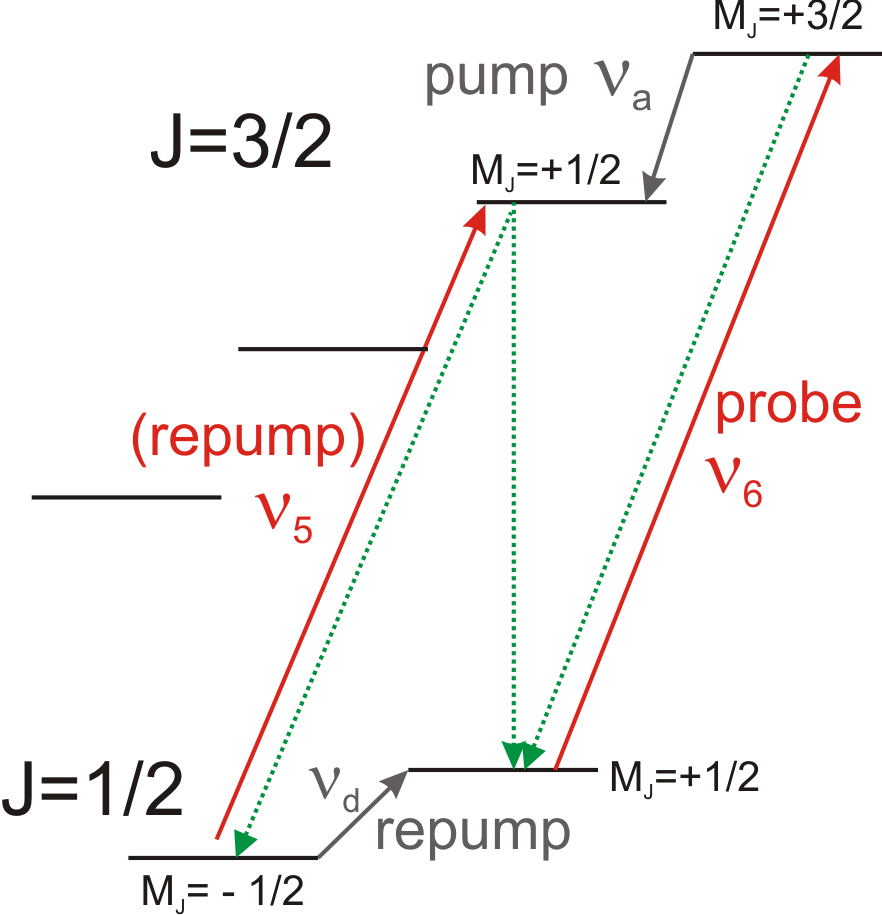}
\caption[Probe spectroscopy]{(Color online)
Probe spectroscopy in the Zeeman-split fine-structure doublet.
Solid arrows indicate the saturated probing and, optionally, pumping laser photons,
while dotted arrows are spontaneous decays.
Gray arrows represent microwave-stimulated transitions.
\label{fig:spec1}}
\end{figure}

The \AngK{1}{+1} population is probed by a laser at frequency $\nu_6$ (see Fig. \ref{fig:spec1}). This
drives the closed transition to the \AngK{3}{+3} sublevel. If the laser
is resonant, we repeatedly see fluorescence photons from roughly half of all ions.
When the microwave field is resonant with the upper-level Larmor frequency, the cycle will be disturbed.
An additional decay channel is opened that leads via \AngK{3}{+1} either back or into the dark state
\AngK{1}{-1}. Therefore, after a pumping period with length depending on the respective intensities and
temporal overlap of the exciting fields, the fluorescence signal will vanish.
A microwave in resonance with $\nu_d$ repumps at least half of the ions (the accurate number again
depends on the relative intensities) back to the bright state. This is a reversible process. We can
measure pumping and repumping efficiencies over and over and arrive at a count rate that depends
on the frequencies only and not on the history of irradiation. This procedure defines a line shape.
However, the scan of a single parameter requires the others to be kept
sufficiently stable. At least the ground-state repumping can be done efficiently by sweeping over the
resonance. This intensifies the signal for finding the upper-level Larmor resonance.
A more general concept is to monitor the yielded fluorescence intensity as a function of all three
frequencies. The multiresonance condition will be represented as a saddle point in this map.

Simultaneous or alternating irradiation with an additional repumping laser beam would facilitate the
measurement. Unfortunately, this is currently not feasible, because the frequencies
$\nu_5$ and $\nu_6$ are separated by 65 GHz. There are no modulators producing such
far-distant sidebands, and a second light source would be needed.
A weaker magnetic field or smaller $g_J$ factor would bring this method back into play.

This method may be inverted in the following sense: The microwave frequency $\nu_a$ is
replaced by $\nu_c$, while the laser frequency is reduced by 325 GHz from $\nu_6$ to
$\nu_1$. Given the
spontaneous transition rate between adjacent Zeeman levels of the order of $10^{-10}$ s$^{-1}$,
the inverse processes are equivalent to the original ones.

\subsubsection{Separation of the cubic effect}
\paragraph{Saturated excitation.}
According to the above arithmetic, many Zeeman coefficients can be derived already from the
frequencies discussed so far.
To improve the coefficient $b_1$ or to get any reliable information about the cubic order,
it is desirable to measure $\nu_b$ as well.
\begin{figure}[h]
\centering
\includegraphics[width=6cm]{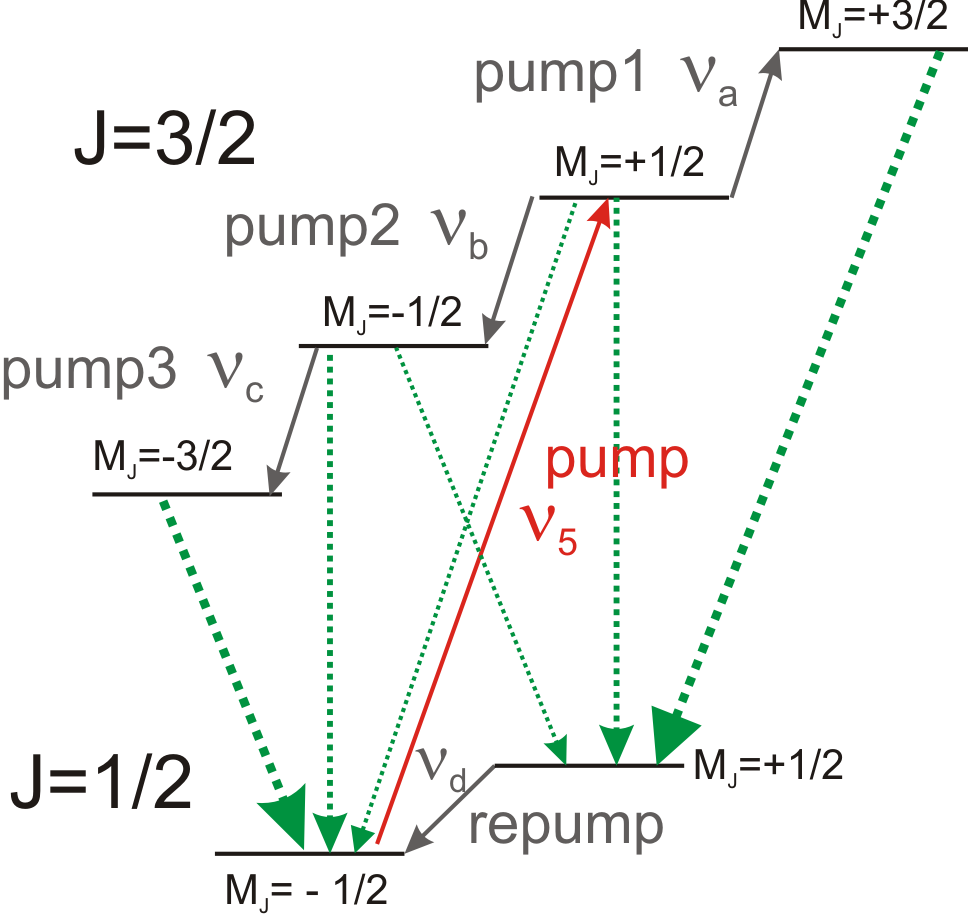}
\caption[Pump spectroscopy]{(Color online)
Pump spectroscopy in the Zeeman-split fine-structure doublet.
Solid arrows indicate the saturated pumping laser photons,
while dotted arrows are spontaneous decays.
Projection-conserving decays of the states \AngK{3}{+1} and \AngK{3}{-1}
are preferred.
Gray arrows represent microwave-stimulated transitions.
\label{fig:spec2}}
\end{figure}
To this end, we look into a method that uses laser pumping instead of probing. This is illustrated
in Fig. \ref{fig:spec2}: For instance,
we depopulate the sublevel \AngK{1}{-1} by exciting ions in this state to the \AngK{3}{+1}
state with the pump frequency $\nu_5$. This level can
decay back to the original state or into the dark state \AngK{1}{+1}. After a few cycles, the
fluorescence will vanish.
Additional irradiation of a microwave at the lower-level resonant frequency will repump ions to the
state \AngK{1}{-1}. A continuous fluorescence signal is a signature of both waves being in resonance
with the corresponding transition.

As a side effect, this would even improve the two-dimensional line shape
(the map of the fluorescence intensity versus
the two frequencies of the laser and the microwave): The maximum of this can
be determined with higher accuracy than the above-mentioned saddle point.
On the downside,
the detectable fluorescence intensity suffers compared to the probe transition,
which is a pure $|\Delta M_J| = 1$ transition with enhanced emission in the axial direction.

This argument leads back to the original chain of thought: We can identify specific substates by the
directional characteristic and branching of different decay modes.
If we drive the upper-level Larmor transition from the \AngK{3}{+1} state to \AngK{3}{+3}, ions
will decay in the projection-changing channel only. This should lead to a more intense optical signal.
However, for this interference-prone signature, it is necessary to have excellent
control of the population distribution in the ground-state sublevels. If the drive works efficiently, we
see yet another effect: The pumping ends faster than before, because \AngK{3}{+1} can
still decay into the bright state.

This study of pumping efficiencies works in a similar way for the transition from \AngK{3}{+1} to
\AngK{3}{-1}. The decay branching of the $\AngK3{+1}$ level is 2:1 in favor of \AngK1{+1}, while the
total decay rate is the same as of the $\AngK3{+3}$ state. Now we can distinguish \AngK3{+1} from \AngK3{-1}
in a spectroscopic experiment: After excitation
from \AngK1{-1} to \AngK3{+1}, ions usually come back with a probability of 1/3, while the remaining ions
fall into the dark state \AngK1{+1}.
A resonant microwave stimulation of the \AngK3{+1} $\leftrightarrow$ \AngK3{-1} transition enhances the
decay into the original state up to a probability of 2/3, which is a factor of 2 in pumping efficiency.
An additional drive of the \AngK3{-1} $\leftrightarrow$ \AngK3{-3} transition could theoretically prevent
the ions
from decaying into the dark state at all.

\paragraph{Coherent excitation.}
\begin{figure}[h]
\centering
\includegraphics[width=6cm]{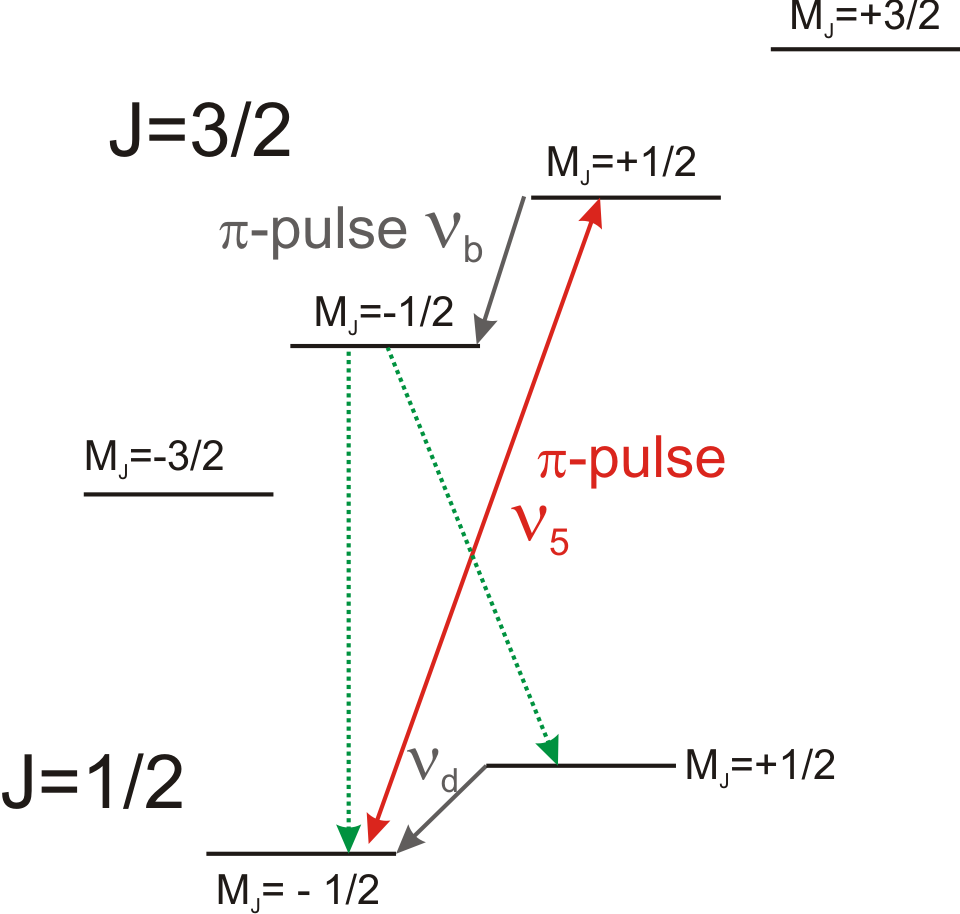}
\caption[Rabi spectroscopy]{(Color online)
Rabi spectroscopy in the Zeeman-split fine-structure doublet.
The solid double arrow indicates coherent excitation or deexcitation by laser photons,
while dotted arrows are spontaneous decays.
Gray arrows represent microwave-stimulated transitions.
\label{fig:spec3}}
\end{figure}
An alternative approach to determine the transition frequency $\nu_b$ can
be based on Rabi spectroscopy \cite{demtroeder},
as shown in Fig. \ref{fig:spec3}.
With laser light of the
frequency $\nu_5$, resonant with the transition \AngK{1}{-1} $\leftrightarrow$
\AngK{3}{+1}, it is possible to drive Rabi oscillations between both
states.
The preparation scheme starts with all population in \AngK{1}{-1}. An
applied laser light pulse with area $\pi$ ($\pi$ pulse) leads to a complete
population transfer to the state \AngK{3}{+1}. From
there, a microwave pulse at a frequency close to $\nu_b$ pumps the
ions to \AngK{3}{-1} with an efficiency depending on power, frequency,
and duration. The remaining population in \AngK{3}{+1} will
be transferred back to the initial \AngK{1}{-1} state with a second
laser light $\pi$ pulse, yielding no detectable light.
In the case of a resonant microwave frequency all particles are in the state
\AngK{3}{-1}, and the last light pulse cannot deexcite the ions.
They will instead decay spontaneously in the subsequent milliseconds
and can with a certain detection efficiency be seen by the detector.
When the frequency of the microwave has been detuned compared to the transition
frequency $\nu_b$, no or less fluorescence will be visible.
After several cycles, the ions will get pumped into the state
\AngK{1}{+1}. Therefore we introduce an additional signature: The population in
this dark state can be monitored with a second laser at frequency $\nu_6$ (probe laser).
This drives the closed transition
\AngK{1}{+1} $\leftrightarrow$ \AngK{3}{+3} and produces fluorescence photons.
A microwave pulse at the lower-level Larmor frequency $\nu_d$ will prepare the initial
state again.

\section{Experimental Setup}
\subsection{Magnet}
\begin{figure}[h!tb]
\begin{center}
\includegraphics[width=0.42\textwidth]{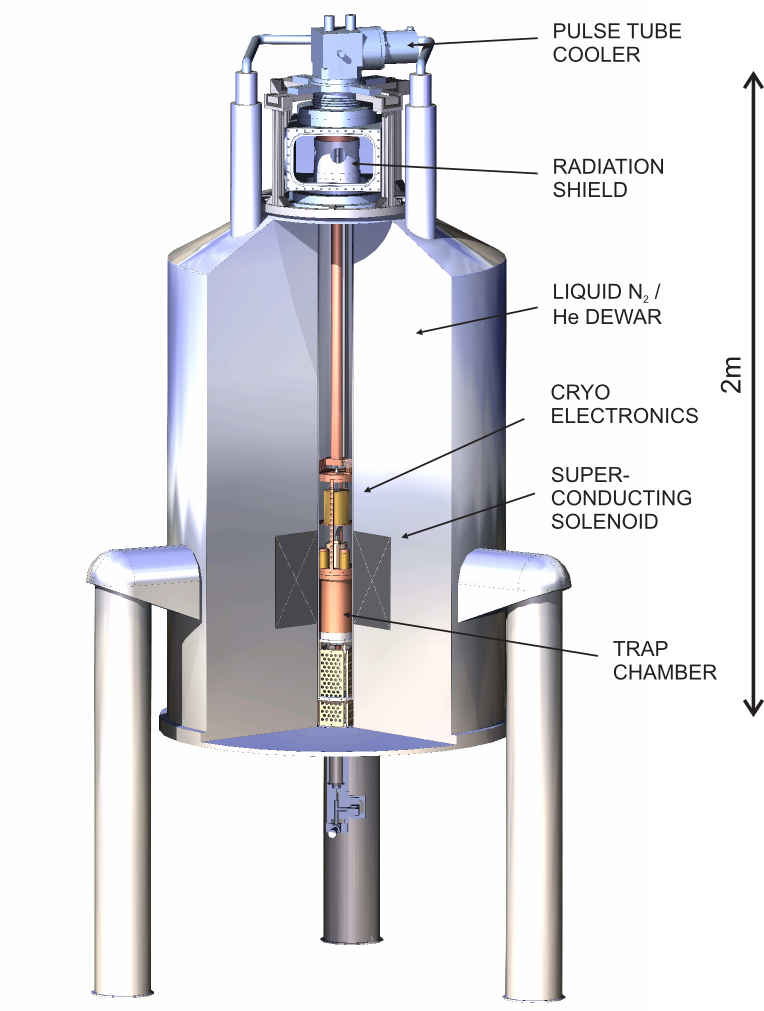}
\caption{\small (Color online) Schematic drawing of the setup. The superconducting magnet has been cut to allow
a view of the trap and related components inside the room-temperature bore.}
\label{magnet}
\end{center}
\end{figure}

The main components of the experimental setup including the Penning trap itself are surrounded by a
vertical room-temperature open-bore magnet (see Fig. \ref{magnet}). It produces the
magnetic field which confines the ions and causes the Zeeman splitting. The maximum field strength is 7
T, and the central field region is located
at the trap center and has a measured homogeneity of 0.14 ppm over the central volume of 1 cm$^3$.

The trap setup is inserted into the 160-mm-diameter magnet bore from the top and represents a shielded
cryostat which employs a pulse-tube cryocooler, to which the Penning trap and the cryoelectronic
components are attached.
The cryostat is a low-vibration pulse-tube cooler with two thermal stages. The first stage has 40 W
cooling power
at 45 K and maintains the temperature of the radiation shield; the second stage inside the radiation
shield has 1 W cooling power at 4.2 K and keeps the trap and its electronics at liquid-helium
temperature. Several aspects of this setup have already been described in \cite{dav,dav2}. The working
principle of in-trap ion production and transfer is similar to that in \cite{haff03}.

\subsection{Trap}
\label{tra}
Penning trap designs and properties as relevant for the present experiment have been discussed in great depth,
for example, in \cite{brown86,gab89,werth,vogpr}. Briefly, in the present experiment, the combination of the
vertical
homogeneous magnetic field of 7 T strength with a harmonic electrostatic potential created by appropriate
voltages applied to the trap electrodes confines ions close to the trap center in all three dimensions.
In the absence of imperfections, each ion performs an oscillatory motion consisting of three
eigenmotions, which can be manipulated individually \cite{dje,sens}. The main manipulation techniques
here are cooling of the ion motion and
compression of the stored ion cloud, as will be discussed below.

\begin{figure}[h!tb]
\begin{center}
\includegraphics[width=0.42\textwidth]{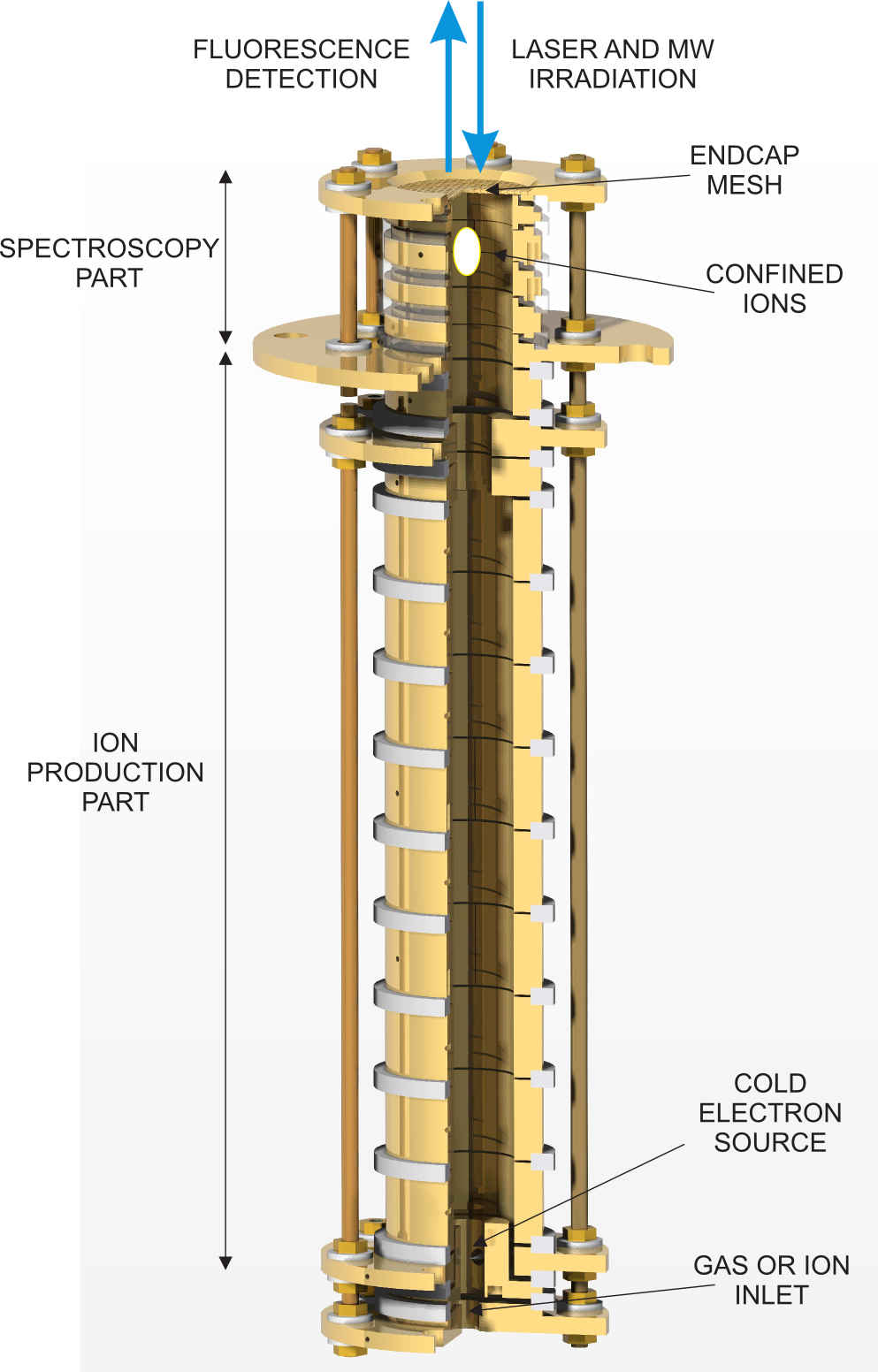}
\caption{\small (Color online) Schematic drawing of the trap. A cut has been applied for presentation of the trap interior.}
\label{trap}
\end{center}
\end{figure}

The trap is a stack of cylindrical electrodes as shown in Fig. \ref{trap}. It consists of a spectroscopy part on the upper end and of an ion production part below. This part features a cold
gas source and a cold electron source (field emission point) and can
be operated like a miniature electron-beam ion source (EBIS) for in-trap production of highly charged gas
ions. To that end,
cold electrons are accelerated towards the spectroscopy trap and deflected back and thus oscillate through the
region where gas
can be injected under well-defined conditions \cite{dav,dav2}. This leads to charge breeding by electron-impact ionization of
gas atoms and ions. The electron energy can be adjusted to any value up to about 2.5 keV, the optimum
production energy
for Ar$^{13+}$ under our conditions has been found to be 1855 eV. To this end, we have performed
simulations using the {\sc cbsim} software package \cite{ dav2}. The evolution of charge states is
monitored by real-time Fourier-transform ion cyclotron resonance mass spectrometry \cite{mar}.
The ion species of interest is selected by resonant ejection of all
unwanted ions \cite{guan} and then transported to the spectroscopy trap by appropriate switching
of electrode voltages.

Alternatively, the trap can be used for dynamic capture and storage of externally produced ions, for
example from an electron-beam ion source or from the HITRAP facility at GSI, Germany \cite{kluge}. In this mode of
operation also, the ions are prepared
in the production part of the trap and then transferred to the spectroscopy part for measurements.
After loading of the trap with ions, cooling of the ion motion is achieved by resistive cooling
\cite{win75} using a tuned circuit attached to the split correction electrode of the spectroscopy
trap. When the resonance frequency of the circuit is tuned to the ion oscillation frequency (in
the first experiments we are mainly interested in cooling of the axial motion), energy from that
ion motion is dissipated into the cryogenic surroundings, thus cooling the motion, i.e., reducing
its energy and therefore its oscillation amplitude. Such resistive cooling to an equivalent of
about 4 K reduces the relative Doppler width of the 441 nm optical transition to about
$2 \cdot 10^{-7}$. The same technique is used for
a nondestructive measurement of ion oscillation frequencies by so-called ``electronic pickup" \cite{win75}.

In the spectroscopy trap, the ions are confined close to the upper end cap such that fluorescence
collection takes place
at a comparatively large solid angle through a partially transparent mesh on top of that electrode.
The excitation laser light
and microwaves are guided into the trap from the top as well.

The ring electrode is split into four equal segments such that a rotating
dipole field can be irradiated into the trap center. This allows the application of the ``rotating wall"
technique for radial compression and shaping \cite{Bha12} of the ion cloud.

\subsection{Laser, microwaves, and detection}
\label{sec:laser}
The optical transitions with frequencies $\nu_1$ to $\nu_6$ of Fig. \ref{fig:spec0} at a wavelength of
441.3 nm will be excited by the radiation of a diode laser (TOPTICA DL 100 pro) with an output power
of 17 mW and an overall tuning range of 6.4 nm.
The manufacturer specifies the spectral width to approximately 1 MHz, which is sufficient to excite
the transitions of interest selectively and is narrow enough to resolve the estimated Doppler broadening
of 150 MHz.
The mode-hop-free tuning range is 22 GHz, easing the search for the transition frequency, which is
known within an uncertainty of about 400 MHz \cite{maeckel:11:prl}. On the other hand, to switch between
two or more of the six transition frequencies, manual tuning of the laser frequencies is required.
Calibration of the laser frequency will be achieved by Doppler-free saturation spectroscopy on molecular
tellurium vapor $^{130}$Te$_2$.
Tellurium has a mapped set of resonance lines in the visible spectrum, especially in the blue.
The ``tellurium atlas'' \cite{atlas} contains a series of lines which are between $2.7$ and 38 GHz
separated from the assumed Ar$^{13+}$ transitions.
As acousto-optic, electro-optic, or sideband modulation might not be sufficient to bridge the
frequency gap to a known tellurium line directly, an offset lock to a second diode laser, which
is stabilized to a tellurium line, can be used for laser frequency stabilization.
Alternatively, a locking scheme based on a frequency-stabilized transfer cavity (see, e.g., \cite{Bilaser})
can be implemented. By a controlled variation of the cavity length, a frequency tuning over the full
mode-hop-free tuning range is accessible \cite{Tellurlines}. In this case, tellurium spectroscopy can
be used for frequency calibration with respect to a known tellurium line. Alternatively, a frequency
comb may be used for improved absolute frequency calibration.

The lifetime of the optical transition shown in Fig. \ref{fig:spec0} at 441.3 nm is 9.57 ms
\cite{lapierre:05:prl}, such that from a single ion a
fluorescence emission rate of about 50 s$^{-1}$ is expected at saturation. The laser intensity needed
for saturation of the closed-cycle transition is about 75 nW\,cm$^{-2}$ for a narrowband laser. With
an assumed linewidth of the excitation laser of about 1 MHz, a laser intensity of a few mW\,cm$^{-2}$
and a total power of about 1 mW are needed to saturate the optical transitions.
This power is readily available from our diode laser system (see above). The required power for the
microwave excitation of the Zeeman
transitions at 65 and 130 GHz is estimated to be of order 10 $\mu$W and is also available
with our 10 mW microwave source, which is frequency stabilized to a 10 MHz external rubidium clock
with a relative
accuracy on the 100 s scale of order 10$^{-12}$.

The overall detection efficiency including solid angle, light transmission of the different optical
elements, and the quantum efficiency of a channel photon multiplier (CPM) detector is estimated to be
about 2$\permil$. Hence, for a cloud of $10^5$ stored ions, the
photon count rate is expected to be of order $10^4$ s$^{-1}$, which allows a good signal-to-noise ratio
for the optical detection.

\subsection{Precision and benefits}
For an evaluation of the measured Zeeman splittings, the magnetic field strength $B$
at the position of the ion has to be determined with high accuracy. This is achieved via the free cyclotron
frequency $\omega_c$ of a single ion with known mass $m$ and use of the relation $\omega_c=qB/m$, where $q$
is the ion charge. Following the ``invariance theorem'' \cite{gab82}, the free cyclotron frequency is given as
the squared sum of all three ion oscillation frequencies. To that end, it is required to measure the reduced
cyclotron frequency $\omega_+$, the axial
oscillation frequency $\omega_z$, and the magnetron drift frequency $\omega_-$ by electronic pickup as
described in detail, for example, in \cite{win75,gru}. For single ions, such measurements have been
performed in numerous variations including
also coupling of individual motions, which reach accuracies of ppb and better \cite{blaum,ulm,stu}.
In principle, on this level of accuracy, measurements are susceptible to charge effects \cite{winters}. However,
for a single ion in our trap, space charge effects are absent and image charge effects can well be corrected for.

The precision of $g_J$ values is then limited by the magnetic field measurement and stability, which
are typically in the ppb regime within the usual measurement times. In the case of
boronlike argon, the upper-level Larmor frequency sets a comparable limit by its natural width.
The Zeeman splitting in longer-lived states, however, can be determined with the significantly higher
accuracy of microwave and rf technology. The present system involves two different
$g_J$ factors. The relation of simultaneously measured Larmor frequencies may profit from this:
In general, one $g_J$ factor may be used as reference for the other instead of the
cyclotron frequency.

This is of particular interest because of the following physical motivation:
The leading-order value of most $g_J$ factors is a rational number given by the Land\'e formula
for coupling of
spin and orbital moments. For instance, in a \textit{P} doublet, the lower and upper levels have $g_{1/2}=2/3$ and
$g_{3/2}=4/3$,
respectively. Deviations from the ratio of exactly 2 are of purely relativistic and
quantum-electrodynamical origin, as discussed in Sec. \ref{sec:theo}.
They usually do not scale with the same rational factor; hence they
are not canceled, but refined from the trivial offset, by the arithmetic operation $g' = g_{3/2} - 2 g_{1/2}$.
This small difference can be measured more directly by modulation of the fundamental microwave oscillation
with a
finite radio frequency instead of generating the radiation with two separate microwave synthesizers.
Carrier and sidebands will be mixed in the up-conversion process and the frequency interval, as defined by
the modulation, is conserved.
Then both Zeeman transitions are driven simultaneously with a single source, and the refined value $g'$
is derived from the radio frequency with its attendant precision. This removes the uncertainty due to
cancellation of two microwave frequencies.

The deviation of this radio frequency from zero reflects the deviation of the actual $g_J$ factors from the
Land\'e formula and allows determination of the anomalous contributions without the precision
restriction caused by the cyclotron frequency.

\section{Summary and Outlook}
We have shown that an understanding of the Zeeman effect at higher orders also is indispensable for
spectroscopy of highly charged ions at the current level of experimental precision. We have provided a
detailed calculation of the first-, second-, and third-order Zeeman effects for a boronlike system as
presently under investigation. Experimental schemes for the separation of the respective contributions
to the Zeeman effect have been given, together with a description of the corresponding experimental setup
for in-trap laser-microwave double-resonance spectroscopy of confined, highly charged ions. Such
measurements yield well-defined access to higher-order Zeeman effects in highly charged ions.

\acknowledgments
This work has been supported in part by DFG (Grants No. VO 1707/1-2 and No. BI
647/4-1) and GSI.
D.A.G., A.V.V., M.M.S., V.M.S., and G.P. received
support from a grant of the President
of the Russian Federation (Grant No. MK-3215.2011.2), by RFBR (Grants
No. 12-02-31803 and No. 10-02-00450), and by the Ministry of Education and Science of Russian Federation. D.A.G.
acknowledges support by the FAIR-Russia Research Center, and by the ``Dynasty'' Foundation. D.L. is supported by IMPRS-QD Heidelberg.


\bibliography{DvL}
\bibliographystyle 
{unsrt}

\end{document}